# Unidirectional Rashba Spin Splitting in Single Layer $WS_{2(1-x)}Se_{2x}$ alloy

Jihene Zribi[1], Debora Pierucci[1], Federico Bisti[2], Biyuan Zheng[3], José Avila[4], Lama Khalil[1], Cyrine Ernandes[1], Julien Chaste[1], Fabrice Oehler[1], Marco Pala[1], Thomas Maroutian[1], Ilka Hermes[5], Emmanuel Lhuillier[6], Anlian Pan[3,||], and Abdelkarim Ouerghi[1,||]

[1] Université Paris-Saclay, CNRS, Centre de Nanosciences et de Nanotechnologies, 91120, Palaiseau,
[2] Dipartimento di Scienze Fisiche e Chimiche, Università dell'Aquila, Via Vetoio 10, 67100 L'Aquila, Italy
[3] Key Laboratory for Micro-Nano Physics and Technology of Hunan Province, State Key Laboratory of Chemo/Biosensing and Chemometrics, College of Materials Science and Engineering, Hunan University, Changsha, Hunan 410082, China
[4] Synchrotron-SOLEIL, Saint-Aubin, BP48, F91192 Gif sur Yvette Cedex, France
[5] Park Systems Europe GmbH. Schildkroetstrasse 15, 68199 Mannheim, Germany
[6] Sorbonne Université, CNRS, Institut des NanoSciences de Paris, INSP, F-75005 Paris, France

Atomically thin two-dimensional (2D) layered semiconductors such as transition metal dichalcogenides (TMDs) have attracted considerable attention due to their tunable band gap, intriguing spin-valley physics, piezoelectric effects and potential device applications. Here we study the electronic properties of a single layer $WS_{1.4}Se_{0.6}$ alloys. The electronic structure of this alloy, explored using angle resolved photoemission spectroscopy, shows a clear valence band structure anisotropy characterized by two paraboloids shifted in one direction of the ***k***-space by a constant in-plane vector. This band splitting is a signature of a unidirectional Rashba spin splitting with a related giant Rashba parameter of $2.8 \pm 0.7$ eV Å. The combination of angle resolved photoemission spectroscopy with piezo force microscopy highlights the link between this giant unidirectional Rashba spin splitting and an in-plane polarization present in the alloy. These peculiar anisotropic properties of the $WS_{1.4}Se_{0.6}$ alloy can be related to local atomic orders induced during the growth process due the different size and electronegativity between S and Se atoms. This distorted crystal structure combined to the observed macroscopic tensile strain, as evidenced by photoluminescence, displays electric dipoles with a strong in-plane component, as shown by piezoelectric microscopy. The interplay between semiconducting properties, in-plane spontaneous polarization and giant out-of-plane Rashba spin-splitting in this two-dimensional material has potential for a wide range of applications in next-generation electronics, piezotronics and spintronics devices.

[||] Corresponding authors E-mail:
abdelkarim.ouerghi@c2n.upsaclay.fr
anlian.pan@hnu.edu.cn

Atomically thin two-dimensional (2D) layered semiconductors such as transition metal dichalcogenides (TMDs) $MX_2$ with M = Mo or W and X = S, Se or Te have attracted considerable attention due to their tunable band gap, intriguing spin-valley physics piezoelectric effects and potential device applications.[1–6] In order to explore the potential properties of these 2D layered materials, their combination into heterostructures can give access to a variety of the chemical, structural and electronic properties of TMDs.[7–9] These heterostructures have been proven to be of great interest for the realization of the next-generation of specific 2D devices, for which the electrical and optical properties are intrinsically determined by the unique energy band alignment at their hetero-interfaces.[10–17] In this regard, TMDs materials are particularly interesting semiconductors as they combine tunable indirect to direct band gap transitions[18] with large exciton binding energies and lack of inversion symmetry[19]. Furthermore, the presence of $d$ orbitals from the heavy transition metal (W, Mo) can lead to strong spin-orbit coupling effects (SOC), which induces an energy splitting of the valence bands at the corners (K points) of the first Brillouin zone (BZ). Due to the lack of the inversion symmetry in monolayer $MX_2$, the spin degeneracy is lifted. Therefore, the bands are spin polarized with opposite polarization in the two K and K' valleys. If the out-of-plane mirror symmetry of the system is also broken, then a Rashba spin splitting can be induced at the Γ point of the BZ. This Rashba spin splitting leads to a shift of the spin polarized bands in the momentum space ($\Delta k_R$) in opposite directions, lifting the spin degeneracy of electronic states around the Γ point[20–23]. The relevant quantities are the magnitude of the Rashba effect induced spin splitting, characterized by the Rashba energy $\Delta E_R$ of the split states, and the Rashba spin splitting parameter[24] $\alpha_R = \frac{2\Delta E_R}{\Delta k_R}$. This Rashba-type spin splitting based on 2D materials has attracted increasing interesting in the field of spintronic applications[25,26]. For instance, the spin field effect - transistor (SFET) as proposed by Datta-Das[27] requires a channel material with large Rashba-type spin splitting bands and an electrically tunable Rashba constant to effectively modulate the electron current by driving the spin precession[28]. Then, various efforts have been made to break out-of-plane mirror symmetry in $MX_2$ monolayers. In particular, recent *ab initio* calculations[22] predicted that this symmetry can be broken for high electric fields (0.1−0.5 V/Å) inducing an anisotropic Rashba spin splitting. The Rashba effect has been demonstrated also in van der Waals heterostructures, where $MX_2$ monolayer have been combined with different materials such as Bi (111)[29] and ferromagnet CoFeB[30]. Recently, ternary compounds of TMDs (MXY, M = Mo or W and X,Y = S, Se, Te where X ≠ Y) have been synthesized in the Janus form, where the top chalcogen X atoms have been completely replaced by Y atoms[31], or in the alloy form[32,33]. Presenting an intrinsic mirror asymmetry, these materials are predicted to display Rasbha effect[34–37] and a possible in plane and/or out of plane electrical polarization[31,38–40]. Despite numerous theoretical studies, experimental evidence of such intrinsic electrical polarization and related Rasbha spin splitting of the bands is still missing.

Here, we use chemical vapor deposition (CVD) technique to prepare monolayer $WS_{1.4}Se_{0.6}$ alloy[41]. Based on theoretical calculations[38], we expect a stable in-plane spontaneous polarization, which we are able to experimentally characterize using piezo-force microscopy (PFM). Angle resolved photoemission spectroscopy (ARPES) measurements show a clear valence band structure anisotropy characterized by two paraboloids shifted in the $\boldsymbol{k}$-space by a constant in-plane vector. This experimental out-of-plane Rashba spin splitting ($\alpha_R = 2.8 \pm 0.7$ eV Å) is ascribed to the in-plane spontaneous polarization observed in the $WS_{2(1-x)}Se_{2x}$ alloy by PFM.

The standard crystal structure of hexagonal TMDs as $MX_2$ monolayer is shown Figure 1(a). Due to the hexagonal symmetry of the monolayer, the in-plane positions of the top and bottom chalcogen sub-layers are superposed. Moving to the band structure, we can describe the effective magnetic field inducing spin precession of carriers as by $\boldsymbol{B(k)} = \alpha(\boldsymbol{E} \times \boldsymbol{k})$, with α a material-dependent parameter accounting for the first-order momentum Rashba spin splitting, $\boldsymbol{E}$ an effective electric field originating from either external potential differences or internal polarization and $\boldsymbol{k}$ the carrier wavevector. Figures 1(b-d) sketch the three basic band spitting types of the valence band at Γ in different spin-orbit field orientations. From left to right, we show the case without field $E$=0 (Figure 1(b)), with a vertical electric field $\boldsymbol{E} = E\hat{\boldsymbol{z}}$ (Figure 1(c)), and with an in-plane electric field $\boldsymbol{E} = E\hat{\boldsymbol{x}}$ (Figure 1(d)). The second and latter case corresponds to the conventional Rashba spin splitting and the out-of-plane Rashba spin splitting, respectively. These two configurations can be easily distinguished in the k-space because the conventional Rashba effect (out-of-plane electric field) induces an isotropic spin-splitting of the bands, while the out-of-plane Rashba effect (in-plane electric field) creates a unidirectional spin-splitting of the bands (i.e., an anisotropic valence band structure modulation) separated by a constant vector $\varDelta k_y$ and a spin-degenerate line at $k_x$=0. The former presents spin textures that are clockwise and anti-clockwise for the two top valence bands, whereas the latter has a persistent spin texture with a k-independent spin orientation (see also Supporting Information Figure S1).

In the present study, our $WS_{2(1-x)}Se_{2x}$ alloy flakes of monolayer thickness are obtained by CVD on $SiO_2$/Si substrate[32]. The results of the growth are shown in Figure 1(e) with a triangular shape (3-fold symmetry) and well-defined W zigzag edges (W-zz)[42–44] (average size ~ 50 μm). In Figure 1(f) and (g), we show the photoluminescence (μ-PL) peak intensity and position mapping images obtained from this typical microscopic triangular flake. The PL peak amplitude, position and linewidth are quite homogenous, suggesting a high crystallinity of the alloy. The only sizeable variations in PL intensity and energy position on the maps relates to the defective edges of the monocrystal (i.e., chalchogens vacancies in the W-zz edges[43]). Small signatures of cracks are also visible (white arrows), starting at the edges and propagating towards the center of the flake following a 3-fold symmetry. We observe that these cracks can induce local small strain variation region and reduce locally the PL intensity, as previously reported[42]. From the PL energy position, we determine the optical band gap of this ternary alloy $E_g$ = 1.88 eV. Then, the composition of the alloy is estimated using a simple mixing law[45] $E_g = xE_{g(WS_2)} + (1 − x)E_{g(WSe_2)}$. The obtained alloy composition $x$ = S/(S+Se) = 0.70 corresponds to $WS_{1.4}Se_{0.6}$. The PL peak position of the edges present a shift of 30 meV toward the higher energies (Figure 1(h), point 1) indicating a higher strain level at the W-zz edges and probably a higher presence of Se or S vacancies[43]. In Figure S2, we show the composition dependent μ-Raman spectra (532nm laser, room-temperature), from three different monolayers ($WSe_2$, $WS_2$ and $WS_{1.4}Se_{0.6}$). For pure $WSe_2$, the $^{WSe2}A_{1g(Se–W)}$ and $^{WSe2}E_{2g(Se–W)}$ modes are superposed at 249.5 cm$^{-1}$ and the typical in-plane TA phonon mode appears at 250.7 cm$^{-1}$. These phonon mode positions are analogous to those reported in a preceding experimental study.[46] For the pure $WS_2$ phase, the $^{WS2}A_{1g(S–W)}$ and $^{WS2}E_{2g(S–W)}$ mode positions are separate at 417.6 cm$^{−1}$ and 350.6 cm$^{−1}$, respectively. The Raman spectra collected from the WSSe single layer is more complex and displays five main modes[32]. The first three are modes from the binaries: $A_{1g(S–W)}$ mode (412.1 cm$^{-1}$), $E_{2g(Se–W)}$ or $A_{1g(Se–W)}$ modes (262.8 cm$^{-1}$), $E_{2g(S–W)}$ mode (324.7 cm$^{-1}$). These are all marginally upshifted from pure $WSe_2$ and slightly downshifted from pure $WS_2$[47]. Next, we find specific modes which are only observed in WSSe alloys[48],

$A_{1g(S-W-Se)}$ (385.6 cm$^{-1}$) and the combined $E_{2g(S-W)} - LA_{(S-W)} + A_{1g(Se-W)} - LA_{(Se-W)}$ mode (159.8 cm$^{-1}$) which mixes LA phonons and Raman modes. The observation of these ternary-specific Raman modes is coherent with the PL result and their relative position with respect to reference data confirms the ternary nature and the composition of the WS$_{1.4}$Se$_{0.6}$ alloy[49]. More information at the atomic scale are given by the high-resolution scanning transmission electron microscopy (HRSTEM) image (Supporting Information Figure S3) where, thanks to the Z-contrast (Z = atomic number), we can clearly recognize the hexagonal organization of the alternative W and S/Se atoms in each unit (green dashed lines). The homogeneous and brightest spots represent the W atoms (Z =74, light blue circle on the image), the S/Se atoms appear darker (Z=16 and Z = 34, respectively). Three level of brightness are present on the image: the dimmest spots representing the S atoms on top of S atoms (S+S atoms, yellow circle), the relative bright spots the S atoms on top of Se atoms (S+Se, red circle) and the even brighter spots the Se atoms on top of Se ones (Se+Se, purple circle). Due to the ternary alloy nature of WS$_{2(1-x)}$Se$_{2x}$ sample, each chalcogen position in the TEM projection (0001) actually consists in the superposition of two atoms, of which each can either be S or Se.

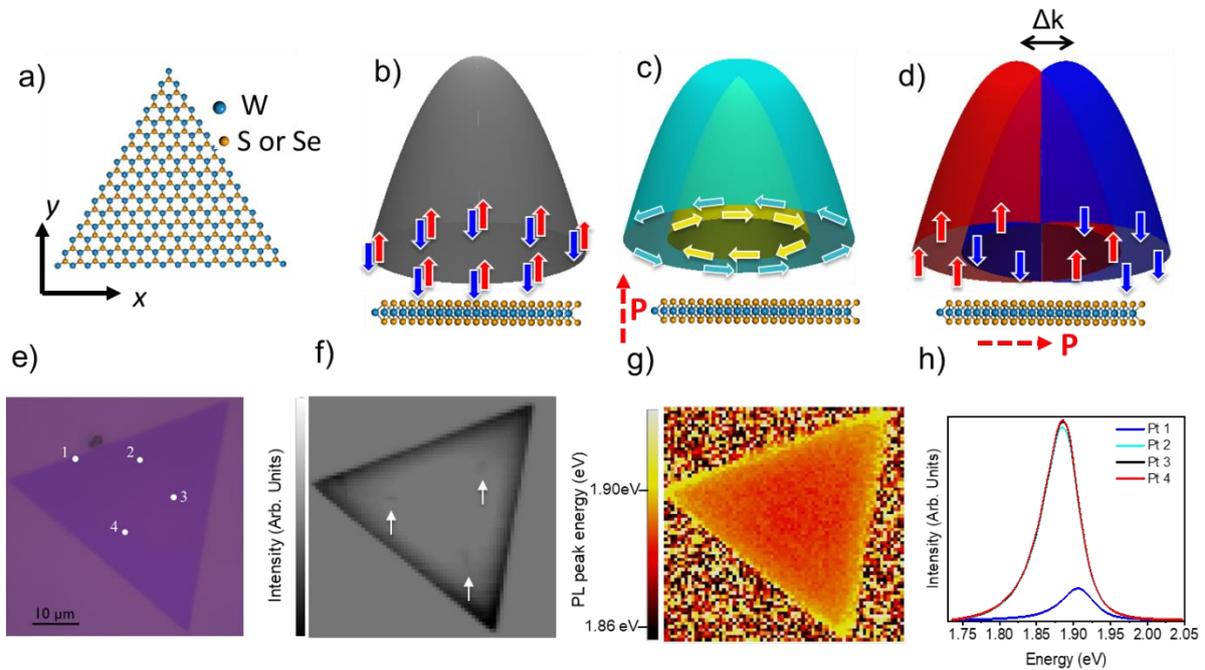

**Figure 1: Structural and optical properties of WS$_{1.4}$Se$_{0.6}$ alloy.** (a) Top view of a monolayer WS$_{1.4}$Se$_{0.6}$ alloy. (b), (c) and (d) Possible band splitting types in momentum space originating in the valence band of a semiconductor from different spin-orbit field orientations. In a two-dimensional electron system, the conventional (c) and out-of-plane Rashba spin splitting (d) appear under the out-of-plane and in-plane polarization, respectively. (e) Optical image, (f) PL intensity map and associated PL peak position map (g). (h) Individual PL (Pt1-Pt4) spectra, acquired at spatial positions (#1-4) in the optical image (e).

Recent calculations have predicted for non-Janus MSeTe and MSSe (M=W or Mo) MLs the presence of a giant in-plane spontaneous electric polarization, with WSSe having the largest one[38]. However, no experimental evidence of such property is present in literature for $WS_{2(1-x)}Se_{2x}$ alloys. To experimentally measure both the electrical polarization and its effect on the band structure, we perform Piezoelectric Force Microscopy (PFM) coupled to angle resolved photoemission spectroscopy with a nanometric x-ray beam at the synchrotron facility (nano-ARPES). To this end, the $WS_{1.4}Se_{0.6}$ flakes were transferred on a conductive graphene/SiC substrate[31] which avoids sample charging effects but also prevent additional stressing or straining on the layer material. Lateral PinPoint™ mode Dual Frequency Resonance Tracking (DFRT) PFM images acquired using a Conductive Diamond Coated Tip are shown in figure 3 (Figure 3(a) topography and 3(b) PFM amplitude). Unlike Janus TDMs[31], $WS_{1.4}Se_{0.6}$ alloy does not present any vertical piezoelectric response. Instead, as shown by figure 3 (b) a clear intrinsic in plane piezoelectric response is present. Combined with the characteristic hysteresis loops observed for both PFM amplitude and phase as a function of applied dc electric field (Supporting Information Figure S5), this response is an signature of a in plane polarization[38]. For parallel comparison, cross-sections along the red lines in both images are compared. The topography shows a relatively homogeneous area with the presence of few wrinkles. In the homogenous parts a clear piezoelectric amplitude of about $180 \pm 10$ µV is present. Note that the PFM amplitude is lower near the wrinkled zone, which proves that the measured in-plane polarization value relates to an intrinsic material property and are not associated to large scale topological defects.

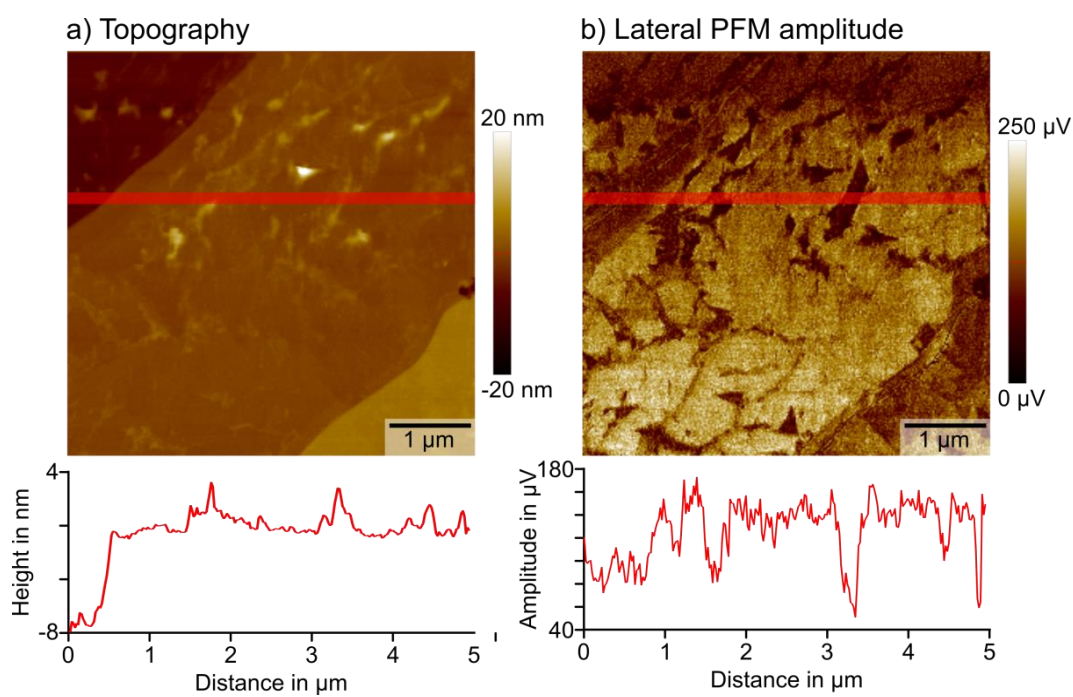

**Figure 2: Characterization of in plane piezoelectricity in the $WS_{1.4}Se_{0.6}$ alloy**: (a) Topography and (b) piezoelectric amplitude (b) of an isolated $WS_{1.4}Se_{0.6}$ monolayer transferred on a graphene/SiC substrate. The layer is uniform in most areas, and shows clear in-plane piezoelectric contrast, except in the area of wrinkles. In the bottom of each image, cross-sections along the red lines are compared at the same tip position.

To directly probe the effect of the measured in-plane polarization on the electronic structure modulations of the $WS_{2(1-x)}Se_{2x}$ alloy, we conduct nano-angle resolved photoemission (nano-ARPES, see Methods) experiments[52,53]. In Figure 3, we compare the electronic structure of a reference binary TMDs monolayer, $WS_2$ (see corresponding PL and Raman data in Supporting Information Figure S4) with that of a ternary $WS_{1.4}Se_{0.6}$ alloy along the ΓK direction (corresponding to W-zz edge direction in the real space, see Supporting Information Figure S6). The sharpness of all the measured bands is indicative of the high crystalline quality of the two samples. For both $WS_2$ and $WS_{1.4}Se_{0.6}$, the top of the valence band at the K point is mostly formed by in-plane $d_{xy}$ and $d_{x2-y2}$ orbital of tungsten, while at the Γ point the valence band is mostly composed by $dz^2$ orbitals of tungsten and $p_z$ orbitals of the chalcogen[54]. For the $WS_2$, the valence band maximum (VBM) at the K point ($E_K^{WS_2}$ =-1.70 eV respect the Fermi level) is higher than the Γ point valence band ($\Delta_{K\Gamma}^{WS_2}$ =-0.20 eV), Figure 3(a)[55,56]. The spin-orbit energy splitting of the valence band at the K point, which mainly originates from the hybridization between the metal $d_{x^2-y^2} + d_{xy}$ and chalchogen $p_x + p_y$ bonding states, is about 420 meV[57–60]. For the $WS_{1.4}Se_{0.6}$ alloy, Figure 3(b), the VBM is also located at the K point, specifically at $E_K^{WSSe}$ =-1.40 eV respects the Fermi level and is $\Delta_{K\Gamma}^{WSSe}$ = -0.41 eV higher than the top of the valence band at the Γ point. We also measure spin-orbit splitting at the K point is 440 meV. From these measurements, we compute, for the $WS_{1.4}Se_{0.6}$ alloy a hole effective mass of 0.5 $m_e$ (upper band) and 0.7 me (lower band), with $m_e$ is the free electron mass at K point. The VBM position at the K point, as underlined by our density functional theory (DFT) calculations (blue dash lines in Figure 3(c-d)) suggests a direct band gap nature of monolayer $WS_{1.4}Se_{0.6}$ alloy[61] as for the standards TMDs materials.

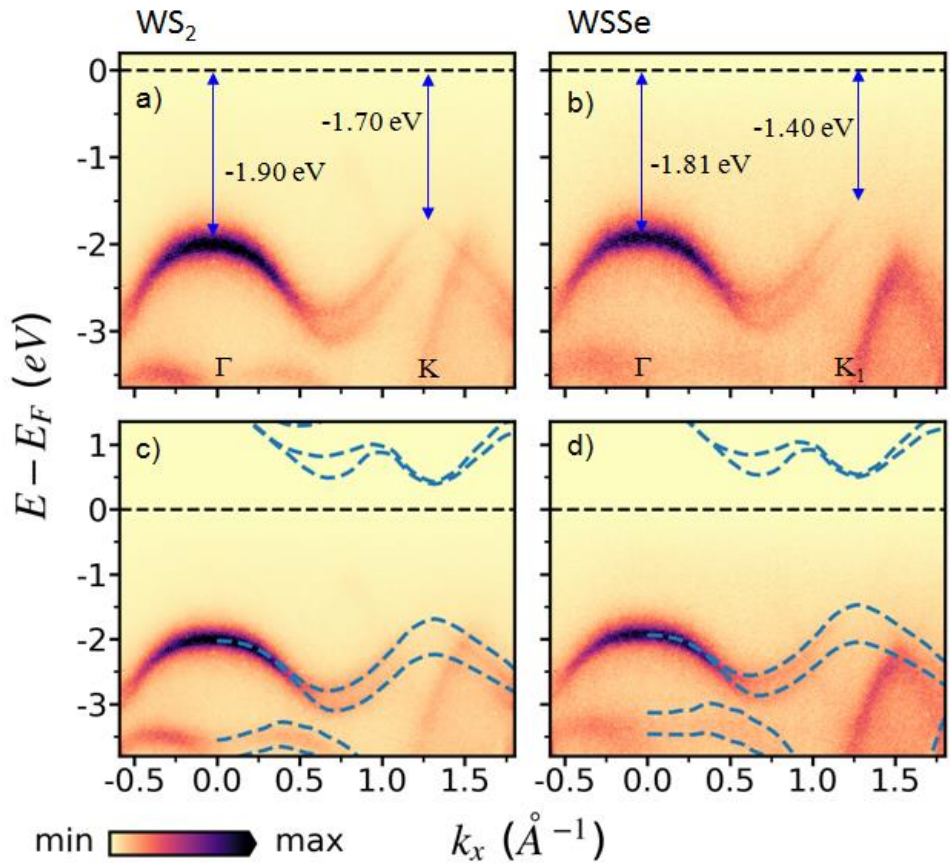

**Figure 3: Comparison between nano-ARPES map of $WS_2$ and $WS_{1.4}Se_{0.6}$ along the ΓK direction:** nano-ARPES map of single layer $WS_2$ (a) and $WS_{1.4}Se_{0.6}$ (b) along the ΓK direction. (c) and (d) shows DFT calculations of valence and conduction bands superposed on the experimental data from parts (a) and (b) as blue dotted lines.

Figure 4 display the ARPES 3D and 2D mapping in $k$-space for the $WS_2$ and $WS_{1.4}Se_{0.6}$, respectively. The $WS_2$ electronic structure (Figure 4(a)) is well described by a single paraboloid centred at the Γ point, and its cuts along the Γ-K (Figures 4(b)) and Γ-M directions (Figures 4(c)) clearly indicates a sharp parabolic profile. On the opposite, the $WS_{1.4}Se_{0.6}$ electronic structure (Figures 4(d-f)) is composed by two paraboloids shifted in one direction of the $k$-space by constant vector ($\Delta k_R$). This vector is found to perfectly align with a specific direction of the BZ (named $M_1$-Γ-$M_1$, see insert panel of Figure 4(d)). This anisotropic electronic structure is of different symmetry respect the $WS_2$. The $C_{3v}$ rotational symmetry is lost and the number of symmetry plane is reduced to a single one, along the direction perpendicular to $\Delta k_R$ (see dashed line in the insert panel of Figure 4(d)) and which contains a specific pair of K and K' points (named as $K_1$ and $K'_1$). Along $K'_1$-Γ-$K_1$ direction (Figure 4(e)) the projection is that of a single parabolic band shape, but actually consists in the superposition of two distinct paraboloids. Along $M_1$-Γ-$M_1$ direction (Figure 4(f)) we now distinguish each of the two parabolic bands, well separated in the $k$-space. An additional representation of these paraboloids is reported in Figure 5, where we show a stack of iso-energy slices, cutting the full tridimensional structure of the band at regular energy intervals. We confirm that only a single paraboloid is observed for $WS_2$, while $WS_{1.4}Se_{0.6}$ exhibits split paraboloids. We assign the observed split in $WS_{1.4}Se_{0.6}$ as a the signature of a unidirectional Rashba spin splitting (see Figure 1(d)), driven by an in-plane electric field directed along the $K'_1$-Γ-$K_1$ direction[62], which breaks the spin degeneracy observed in $WS_2$. In turn, the existence of such electric field can be ascribed to a unidirectional in-plane polarization present in the $WS_{1.4}Se_{0.6}$ alloy.. This peculiar field orientation could be related to possible local ordered configuration present in the $WS_{1.4}Se_{0.6}$ alloy. From the point of view of application, the observed Rashba spin splitting is described by a momentum offset ($\Delta k_R$) away from the crossing point of 0.095 Å and a corresponding energy offset ($\Delta E_R$) of 0.15 eV (see Figure 4(f) and Supporting Information Figure S7). Our analysis shows that the out-of-plane Rashba spin splitting is due to a large in-plane spontaneous polarization in $WS_{2(1-x)}Se_{2x}$ alloy with a Rashba parameter of 2.8± 0.7 eV Å which are among the first giant Rashba spin splitting size parameters reported in the single layer of 2D materials. This Rashba parameter is high in comparison with the theoretically reported for janus WSSe monolayers, $\alpha_R \sim 0.005$ eV·Å[63,64]". However, other recent calculations have shown that such Rashba parameters computed through DFT may vary by one or two order of magnitude with extremely small strain[36]. This Rashba parameter even compares favorably to standard InGaAs/GaAs quantum dots, $\alpha_R \sim 0.08 - 0.12$ eV·Å,[65] heavy metal films including the Au(111) surface ($\alpha_R = 0.33$ eV·Å)[66] and Au/W(110) quantum well ($\alpha_R = 0.16$ eV·Å)[67]. Our 2D Giant Rashba effect is also same order than the recently reported for bulk BiTeI[24] and of the same order of the giant Rashba spin splitting observed in bulk α-GeTe[68]. Such a remarkably large Rashba parameter can significantly improve the spin-charge conversion in spintronic devices[69,70].

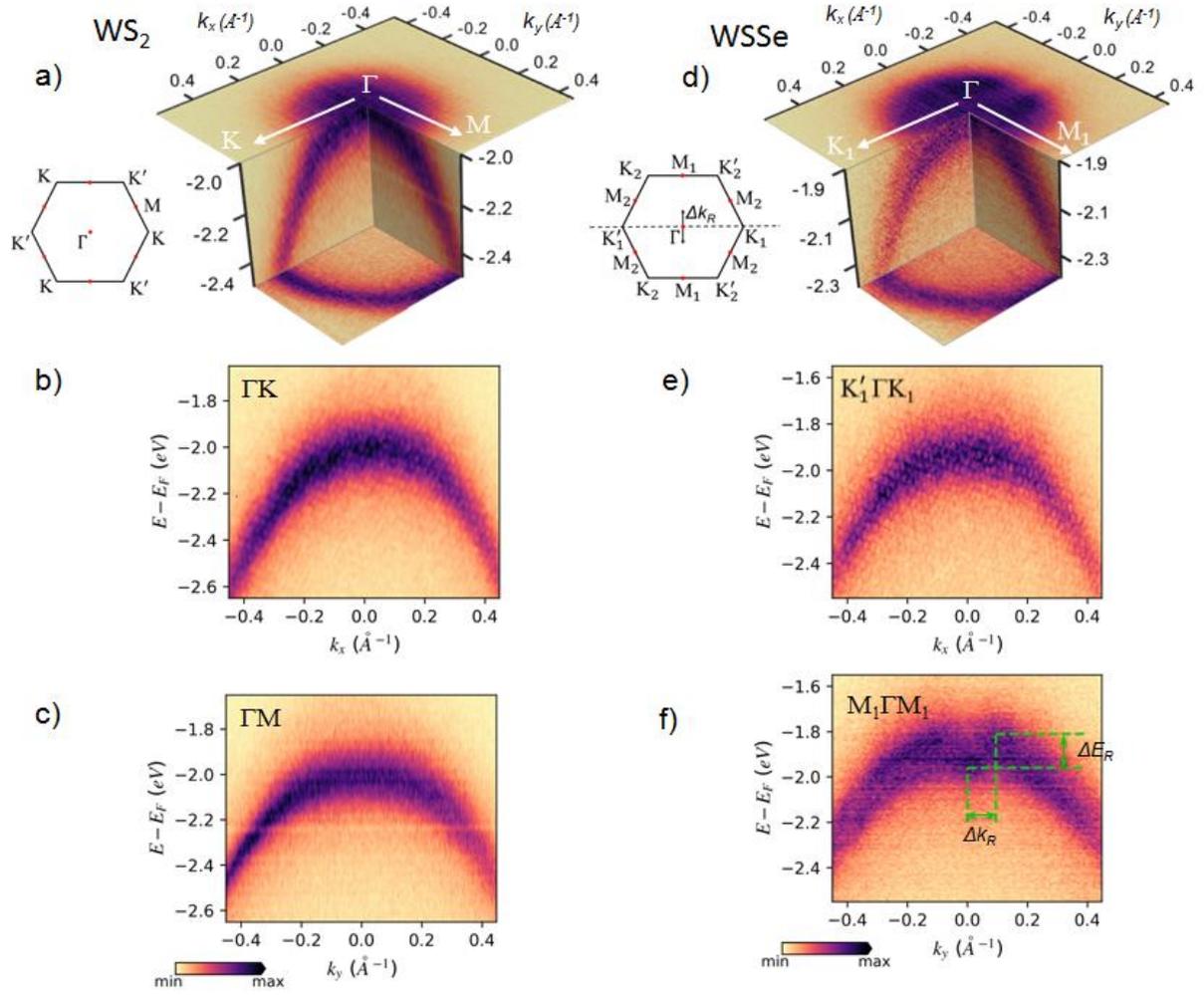

**Figure 4: Comparison between nano-ARPES map of $WS_2$ and $WS_{1.4}Se_{0.6}$ at the $\Gamma$ point**: (a) ARPES 3D mapping in *k*-space and the related BZ in the insert, cuts along (b) $\Gamma$-K and (c) $\Gamma$-M high symmetry directions for $WS_2$. (d) ARPES 3D mapping in *k*-space and the related BZ in the insert, cuts along (e) $K'_1$- $\Gamma$-$K_1$ and (f) $M_1$- $\Gamma$-$M_1$ high symmetry directions for $WS_{1.4}Se_{0.6}$.

We briefly discuss the origin of the in-plan polarization. Standard hexagonal $WS_2$ monolayers obeys the 3m point group symmetry, which gives rise to non-null elements of the piezoelectric tensor due to the lack of symmetry along the armchair directions. However, the piezoelectric tensor coefficients are zero for the zigzag directions (i.e. parallel to zigzag edges of the crystal structure) due to existing inversion symmetry along these lines. As the $\Gamma$-K directions in reciprocal space align with the zigzag directions in real space, the zero coefficients of the piezoelectric sensor impose an absence of polarization along the $\Gamma$-K directions of the $WS_2$ band structure. In the ternary alloy WSSe monolayer, this ideal symmetric configuration is disrupted by the difference in atomic sizes and electronegativities between S and Se atoms. In an idealized TMD ternary random alloy, the chemical nature of the chalcogen atoms is fully independent of its neighbors. Still, the replacement of one site of chalcogen (say S) with another chalcogen atom (say Se) of different atomic radii, will change the local distances to the hexagonal W sub-lattice sandwiched in between the S and Se layers and break down the symmetry of the binary TMDs system. This local symmetry breaking gives rise to finite piezoelectric coefficients along arbitrary orientations. The strain relaxation of such randomly distributed ternary alloy thus gives rise to a distorted crystal structure, in which in-plane electric fields can appear and generate the unidirectional Rashba SOC (see Supporting Information section VII). As these dipoles may exist anywhere inside the WS$_{1.4}$Se$_{0.6}$ crystal, including the previously field-free the zig-zag (real space) or Γ-K (reciprocal space) directions, it is possible to generate the observed unidirectional Rashba SOC along Γ-M.

In practice, real ternary material often exhibits a more complex structure in which chemical fluctuations are spatially correlated. Several studies concerning TMDs alloys[71–73] have demonstrated that fluctuations along the growth direction can lead to ordered configurations in the form of atomic chains oriented parallel to the flake edges (i.e. zigzag directions), akin to ordering observed in the epitaxy of 3D bulk crystals[74]. These chemically ordered configurations of the 2D alloy are mostly generated by the interplay between local lattice distortion and long-range strain interactions. Similarly to the idealized random alloy, the different sizes of the S and Se atoms, and their different electronegativity, will create macroscopic strain (as shown by PL) and local electric dipoles (as shown by PFM). However due to 2D ordering in the ternary alloy (i.e. spatial chemical correlations), the in-plane component of the electrical field may be stronger in particular crystal directions. Here, the local in-plane electric field aligns along a Γ-K direction, i.e. along the zigzag edges, which is perpendicular to the expected growth direction. Ordering in the ternary alloy may thus link the final crystal shape to the in-plane electrical dipole orientation and the particular Γ-M orientation of the unidirectional Rashba SOC observed in our sample. In summary, the alloying process first allows the presence of an in-plane field, while its ordering may favor particular crystal orientation, resulting in the observed Rashba SOC along the specific Γ-M direction. The presence of this dipole provides a 2D platform to study light–matter interactions where dipole orientation is critical, such as dipole–dipole correlations and strong coupling with plasmonic structures.

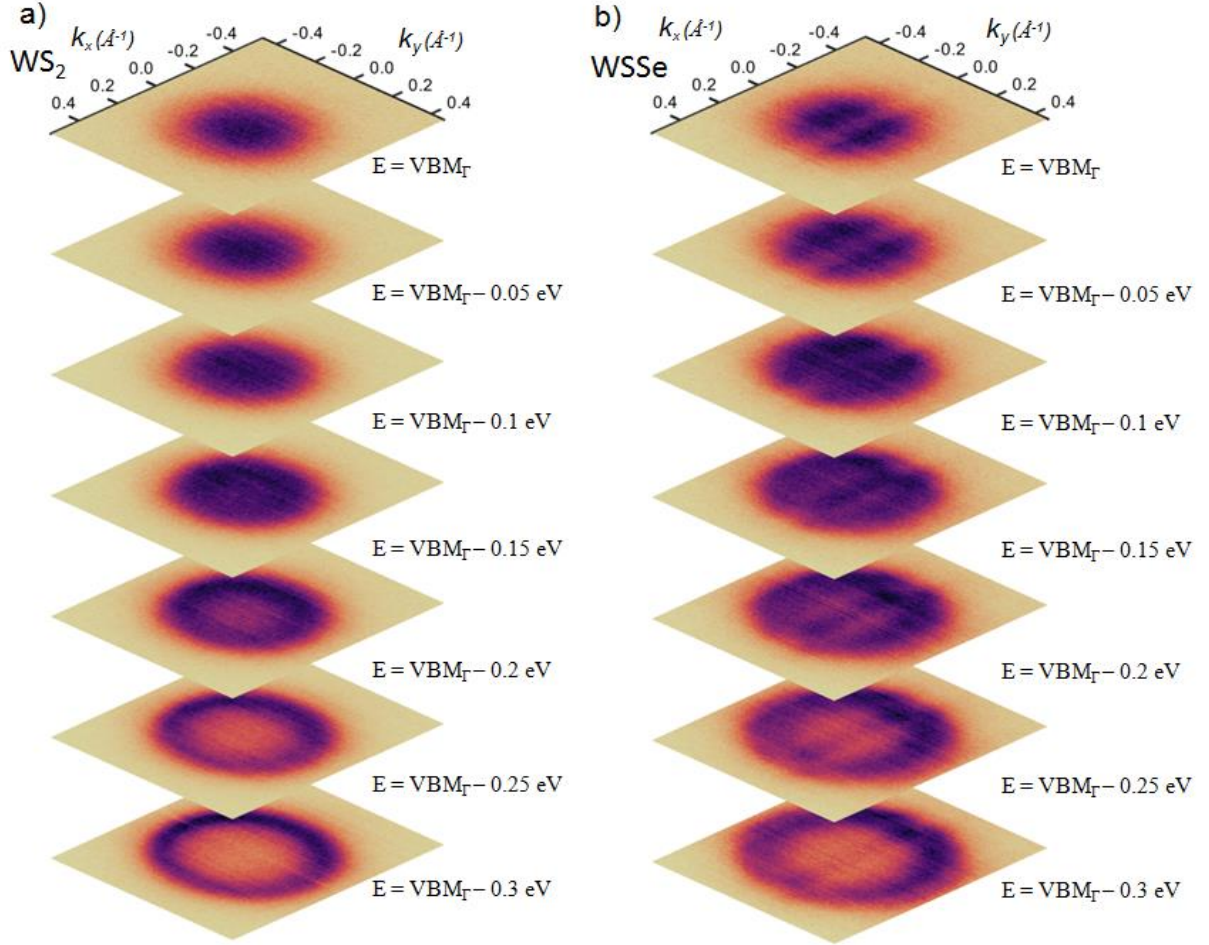

**Figure 5: Comparison between 3D nano-ARPES map of WS$_2$ and WS$_{1.4}$Se$_{0.6}$:** Iso-energy contours of WS$_2$ (a) and WS$_{1.4}$Se$_{0.6}$ (b) of the VBM at the Γ point and different cuts below this point.

In conclusion, we investigate the electronic properties of WS$_{2(1-x)}$Se$_{2x}$ monocrystals of monolayer thickness. We find that WS$_{1.4}$Se$_{0.6}$ gives rise to sharp and well-defined bands showing a giant out-of-plane Rashba effect. Considering that the monolayer WS$_{1.4}$Se$_{0.6}$ is stable in air, despite being only a monolayer thick, it can seamlessly assembled with different rotation angles with other 2D material of different kind allowing an easily band engineering through tunable moiré super lattices[53] and strain[75,76]. This band engineering could allow the valence band maximum of the system to be moved from the K point to the Γ point, in order to obtain well isolated Rashba spin splitting state. Moreover, due to the semiconducting nature of the ternary alloy, the Fermi level can be easily tuned to cross this bands *via* doping[77] or a gate, providing exciting opportunities to fabricate electrically controllable spintronics devices. We expect that the present observations of a giant out-of-plane Rashba spin splitting induced by an in-plane electrical polarization can also be achieved in other 2D alloyed TMDs, which share similar structural and electronic properties. The coexistence of unidirectional Rashba spin splitting (out-plane spin direction) and band splitting at the K and K′ valleys (out-of-plane spin direction) also makes WSSe interesting for valleytronics. Thus, our findings open up new avenues for exploring exotic electrical and optical phenomena in 2D materials systems, as well as novel spintronic devices driven by the Rashba effect.

**Methods**

**Growth of $WS_{1.4}Se_{0.6}$:** The $WS_{1.4}Se_{0.6}$ /SiO$_2$/Si monocrystals were grown through a one-step chemical vapor deposition (CVD) process[41]. At the beginning, 1000 SCCM pure Ar gas was introduced into the system for 10 minutes to ensure a stable chemical reaction environment. Then, the center temperature of the furnace was heated to 1100 °C for 10 minutes and the rate of Ar gas flow was set to 100 SCCM to obtain $WS_{1.4}Se_{0.6}$ monocrystals. The as-prepared sample were transferred onto the graphene/SiC. Before any measurement, the samples were annealed at 250 °C in ultrahigh vacuum, in order to remove any residual surface contaminations left by the wet [78].

**PFM measurement:** Lateral dual frequency resonance tracking (DFRT) piezoelectric force microscopy (PFM) in PinPoint™ mode was measured on a NX10 AFM (Park Systems, Suwon, Korea) with a Zurich Instruments (Zurich, Switzerland) HF2 lock-in amplifier (LIA) via a conductive diamond-doped CDT-CONTR cantilever (Nanosensors, Neuchatel, Switzerland). PinPoint™ mode was used for topography imaging by performing fast force spectroscopy curves (30 µm/s for approach and retract) at each pixel. Once the force setpoint is reached at the lower turning point of the spectroscopy curve, the average lateral PFM signal was collected over a period of 10 ms. For DFRT PFM, an AC sample voltage was applied via the Zurich Instruments HF2 LIA and swept over a frequency range of 300 -500 kHz to identify the lateral contact resonance at ~380 kHz. The two sidebands for the frequency tracking were generated at ±2 kHz with 2 V AC excitation. During the scan, the DFRT feedback kept the amplitude ratio of both sidebands constant by readjusting the AC frequency to match the lateral contact resonance. The DFRT PFM signals were fed back into the atomic force microscopy (AFM) controller using the auxiliary outputs of the HF2 LIA and the auxiliary inputs on the AFM controller.

**Photoluminescence and Nano-ARPES measurements:** Photoluminescence (PL) measurements were conducted using a commercial confocal Horiba micro-Raman microscope with an objective focused 532 nm laser in an ambient environment at room temperature[41]. The Nano-ARPES measurements were performed at the ANTARES beamline of Synchrotron SOLEIL. The ARPES data were taken at a photon energy of 100 eV, spot size of about 500 nm, using linearly polarized light at a base pressure of $5 \times 10^{-11}$ mbar[78].

**DFT simulations:** Band structures calculations were obtained by means of the QUANTUM EPSRESSO suite[79]. We used a fully relativistic pseudopotential and non-collinear simulations to consider the spin-orbit interaction. For the exchange-correlation term we considered the HSE hybrid functional[80] to better estimate the band-gap energy. The self-consistent solution was obtained by adopting a 15x15x1 Monkhorst-Pack grid and a cutoff energy of 50 Ry. A vacuum space of 20 Å along the vertical direction was used to minimize the interaction between two adjacent sheets. Cell parameters and atomic positions were relaxed according to a convergence threshold for forces and energy of $10^{-3}$ and $10^{-4}$ (a.u.), respectively. The $WS_{1.4}Se_{0.6}$ alloy was modeled within the virtual crystal approximation by using a chalcogen pseudopotential given by a linear interpolation of the pseudopotentials for S and Se. The use of the VCA in the context of plane-wave DFT using pseudopotentials is justified when the atoms composing the alloy are similar and when the pseudopotentials use the same number of valence orbitals for each angular momentum, as in the case of S and Se. Recently, the VCA method has been successfully employed in simulating the electronic properties of transition metal chalcogenide alloys[81,82].

**Data availability:** The data sets generated during and/or analyzed during the current study are available from the corresponding author on reasonable request.

**Competing interests:** The authors declare no competing interests.

**Contributions:** B.Z. and A.P. fabricated the samples and carried TEM measurement. J.Z. D.P., A.O., and J.A., carried out the nano-XPS/nano-ARPES experiments. J.C., C.E., L.K., characterized the samples by means of μ-Raman/PL spectroscopy and analyzed the Raman/PL data. I. H, carried the PFM measurement. M.P. carried the DFT calculation. J.Z., F.B., J.A., D.P., F.O., A.O., and E.L. analyzed the data. All the authors discussed the results and commented on the manuscript.

**ACKNOWLEDGMENTS:** We acknowledge the financial support by MagicValley (ANR-18-CE24-0007), Graskop (ANR-19-CE09-0026), 2D-on-Demand (ANR-20-CE09-0026), and MixDferro (ANR-21-CE09-0029 grants, the French technological network RENATECH, the National Natural Science Foundation of China (Nos. U19A2090 and 62090035), the China Postdoctoral Science Foundation (No. 2020M680112), and the Science and Technology Innovation Program of Hunan Province (No. 2020RC2028).

# Unidirectional Rashba Spin Splitting in Single Layer WS$_{2(1-x)}$Se$_{2x}$ alloy

Jihene Zribi[1], Debora Pierucci[1], Federico Bisti[2], Biyuan Zheng[3], José Avila[4], Lama Khalil[1], Cyrine Ernandes[1], Julien Chaste[1], Fabrice Oehler[1], Marco Pala[1], Thomas Maroutian[1], Ilka Hermes[5], Emmanuel Lhuillier[6], Anlian Pan[3,∥], and Abdelkarim Ouerghi[1,∥]

[1] Université Paris-Saclay, CNRS, Centre de Nanosciences et de Nanotechnologies, 91120, Palaiseau,
[2] Dipartimento di Scienze Fisiche e Chimiche, Università dell'Aquila, Via Vetoio 10, 67100 L'Aquila, Italy
[3] Key Laboratory for Micro-Nano Physics and Technology of Hunan Province, State Key Laboratory of Chemo/Biosensing and Chemometrics, College of Materials Science and Engineering, Hunan University, Changsha, Hunan 410082, China
[4] Synchrotron-SOLEIL, Saint-Aubin, BP48, F91192 Gif sur Yvette Cedex, France
[5] Park Systems Europe GmbH. Schildkroetstrasse 15, 68199 Mannheim, Germany
[6] Sorbonne Université, CNRS, Institut des NanoSciences de Paris, INSP, F-75005 Paris, France

∥Corresponding authors E-mail:
abdelkarim.ouerghi@c2n.upsaclay.fr
anlian.pan@hnu.edu.cn

## I. Two-dimensional electron system under in-plane and out-of-plane polarization:

The emergence of either the conventional in-plane or the unidirectional out-of-plane Rashba SOC can be qualitatively evidenced by measuring the shifting direction of the parabolic bands in the k-space. Its spin-texture depends on the orientation of the spin-orbit field

$$\vec{\Omega}_{SOF}(\vec{k}) = \alpha(\hat{e} \times \vec{k})$$

which determines the effective spin-orbit coupling term $H_{SOC} = \vec{\Omega}_{SOF}(\vec{k}) \cdot \vec{\sigma}$, where $\hat{e}$ is the direction of the electric field and $\vec{\sigma}$ is the Pauli spin matrices.

In the conventional spin-orbit Rashba effect, where the electric field is along the *z*-axis, the effective spin-orbit field resides in the *x-y* plane and reads $\vec{\Omega}_{in-plane}(\vec{k}) = \alpha(\hat{z} \times \vec{k})$. Hence, the parabolic bands undergo a circular spin-splitting so that any cross section ($k_z$,$k_{xy}$) will be identical (see Fig. S1(b)). Instead, in the unidirectional Rashba effect, the effective electric field resides in the *x-y* plane and the spin-orbit field $\vec{\Omega}_{out-of-plane}(\vec{k}) = \alpha(\hat{x} \times \vec{k})$ is oriented along the vertical direction (*z*-axis). Thus, the spin texture is identical to systems with equivalent Dresselhaus and Rashba SOC, as shown in Fig. S1(c). The spin-up and spin-down bands are shifted by a constant $\Delta k_y$ along the direction orthogonal to the in-plane polarization (*x*-axis). Therefore, the cross section taken at $k_y=0$ will not show any spin-splitting, whereas the one at $k_x=0$ will present the maximal spin splitting.

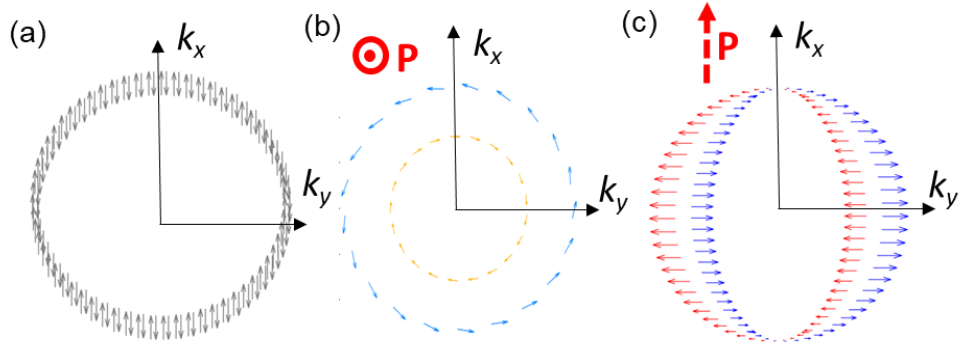

**Figure S1:** In a two-dimensional electron system with parabolic band profiles (a), the conventional (b) and (c) out-of-plane Rashba spin–orbit couplings appear in the presence of out-of-plane ($z$-axis orientated) and in-plane ($x$-axis oriented) polarization, respectively. The conventional Rashba SOC induces a circular spin polarization, while the out-of-plane Rashba SOC shifts the spin-up and spin-down parabolic bands by a constant vector. The (c) case has spin quantization axis along the $z$ axis.

## II.      Micro-Raman spectroscopy of WSSe

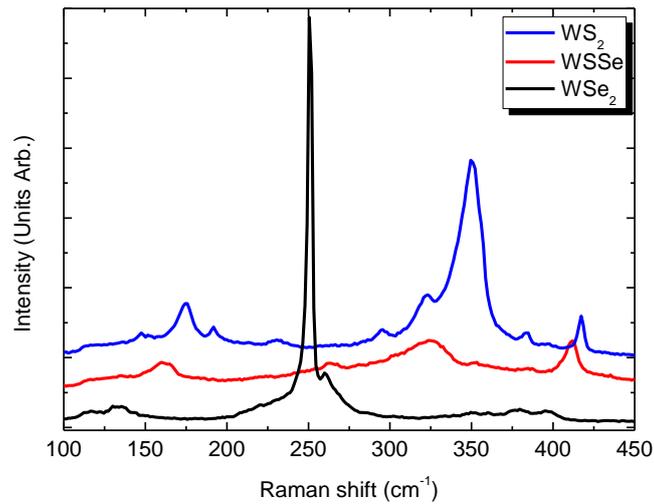

**Figure S2:** Room temperature micro-Raman spectroscopy (532nm) of $WSe_2$, $WS_2$ and $WS_{1.4}Se_{0.6}$ monolayers.

### III. Structural properties of WS$_{1.4}$Se$_{0.6}$ monolayer flakes HRSTEM analysis:

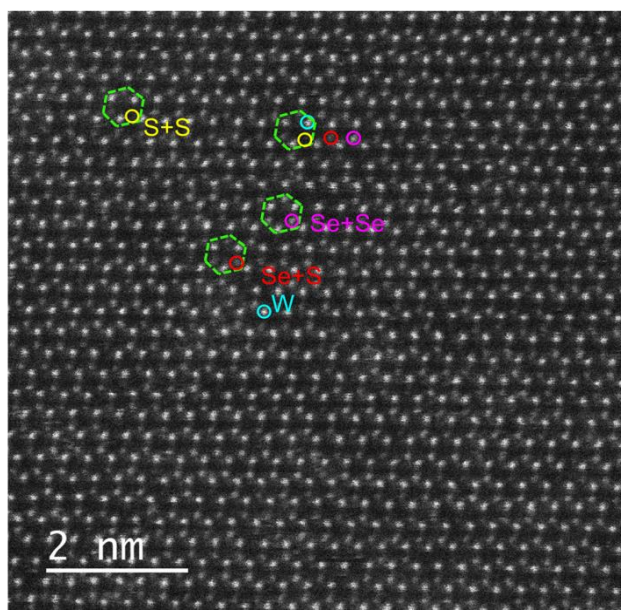

**Figure S3:** HRSTEM image of the WS$_{1.4}$Se$_{0.6}$ monolayer. The hexagonal rings (green dotted lines) consist of alternative W and S/Se atoms. The spots with different brightness in the hexagonal rings represent the locations of W atoms (light blue circle) Se + Se atoms (purple circle), Se + S atoms (red circle), and S + S atoms (yellow circle).

### IV. WS$_2$ monolayer flakes:

WS$_2$/SiO$_2$ samples were grown by chemical vapor deposition (CVD) in a 1" quartz tube furnace. The growth substrate was placed in the center of the furnace and heated to 800 °C. A 25 mg sulfur pellet was placed on a piece of silicon and positioned upstream in the furnace such that its temperature was approximately 150 °C. Carrier gas (500 sccm N$_2$) was used to bring sulfur vapor into the furnace for a 30 min growth period. The CVD growth of WS$_2$ on SiO$_2$ results in characteristic single-crystal domains shaped as well-defined equilateral triangles with a lateral size of about about 50 μm as shown in the optical image of Figure S4(a).

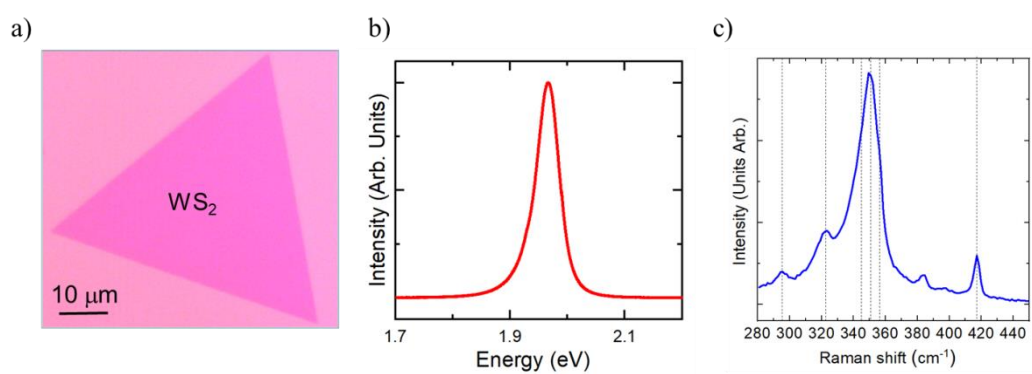

**Figure S4:** (a) optical image, (b and c) PL and Raman spectra of single layer WS$_2$

To investigate the structural properties of the WS$_2$ flakes, photoluminescence (μ-PL) and micro-Raman spectroscopy were used (Figure S4 (b) and S4(c), respectively). The PL spectrum present a sharp peak located at 1.97 eV[1]. In the Raman spectrum beside the first order modes at the Brillouin zone (BZ) center (Γ), the in-plane phonon mode $E_{2g}^1$ at 356 cm$^{-1}$ and the out-of-plane phonon mode A$_{1g}$ at 418 cm$^{-1}$,[2,3] a series of overtone and combination peaks are presents. In particular, the Raman feature around 350 cm$^{-1}$ is the convolution of several components: the $E_{2g}^1$ (Γ), the 2 LA (M) mode at 351.7 cm$^{-1}$, which is a second-order Raman mode due to LA phonons at the M point of the BZ zone, and the $E_{2g}^1$ (M) mode[4]. Moreover, Raman peaks at 323.7 cm$^{-1}$ and 297.1 cm$^{-1}$ are combinations modes, which are attributed to $2LA\ (M) - E_{2g}^2\ (\Gamma)$ and the $2LA\ (M) - 2E_{2g}^2\ (\Gamma)$ modes, respectively.

## V.      Piezoelectric Force Microscopy (PFM) on WS$_{1.4}$Se$_{0.6}$

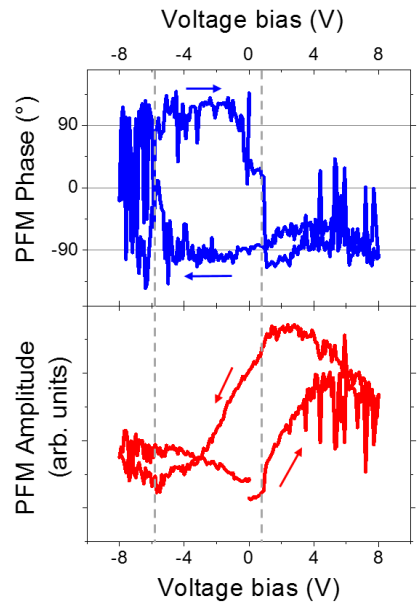

**Figure S5:** Hysteresis loops as a function of applied dc voltage of the lateral piezoresponse phase (top, in blue) and amplitude (bottom, in red). Arrows indicate the direction of the voltage sweep for each branch. Lateral (in-plane) PFM data acquired in dual frequency resonance tracking (DFRT) mode, with 3 Vac applied on each sideband at +/- 2 kHz from the lateral contact resonance.

## VI. Unit cell and corresponding reciprocal space:

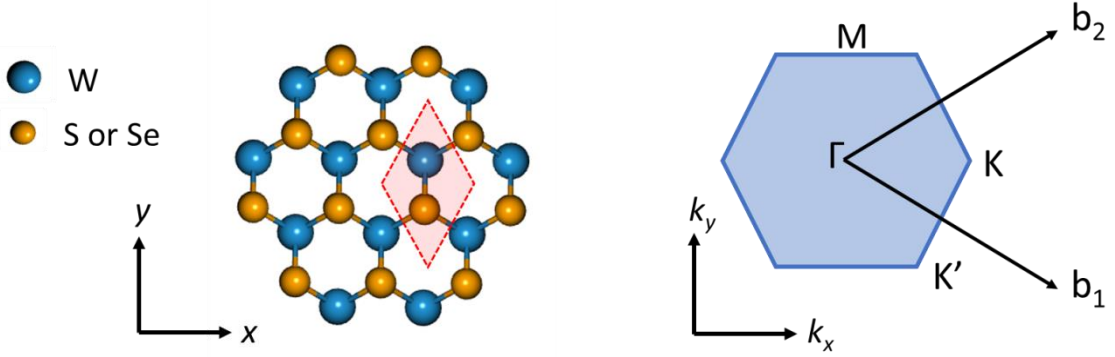

**Figure S6:** unit cell (left) and corresponding Brillouin zone (right) of the $WS_{2(1-x)}Se_{2x}$ crystal. The zig-zag (arm-chair) edges are oriented along the ΓK (ΓM) directions.

## VII. Determination of the Rashba parameter $α_R$ :

To determine the Rashba parameter we adopted a self-consistent method illustrated in Figure S7. From the band dispersion map along the Γ-M direction (top left panel) we extract the momentum distribution curves (MDC) at -1.81 eV integrated over ± 0.01 eV, [blue region in Figure S7(a-b) and blue data and line Figure S7(c)] and at -1.95 eV integrated over ± 0.01 eV, [blue region in Figure S7(a, b) and blue data and line in Figure S7(c)], which correspond respectively to the valence band energy maximum and its minimum at the Γ point. The Energy distribution curves (EDC) are extracted at Γ and at the maximum of the MDC at -1.81 eV, i.e. ky = -0.1, 0, 0.1 Å$^{-1}$ integrated over ± 0.01 Å$^{-1}$ (i.e., $\Delta k_R$ = 0.10 ± 0.01 Å$^{-1}$) [green, red and magenta regions in Figure S7(a, c) and related data and lines in Figure S7(b), respectively]. The $\Delta E_R$ value is obtained by considering the maximum and minimum of the difference between the EDC at Γ (red curve) and the other two EDCs (violet and green curves). For making the difference between these curves, the data had been filtered using a Gaussian filter. The maximum and minimum are located at -1.81 eV (which is then self-consistent with the MDC cut) and -1.95 eV, respectively. Therefore, the obtained $\Delta E_R$ value is of 0.14 ± 0.02 eV, giving a Rashba parameter of $α_R = \frac{2\Delta E_R}{\Delta k_R}$ = 2.8 ± 0.7 eV Å. The error propagation has been derived as $\frac{\delta α_R}{α_R} = \frac{\delta \Delta E_R}{\Delta E_R} + \frac{\delta \Delta k_R}{\Delta k_R}$.

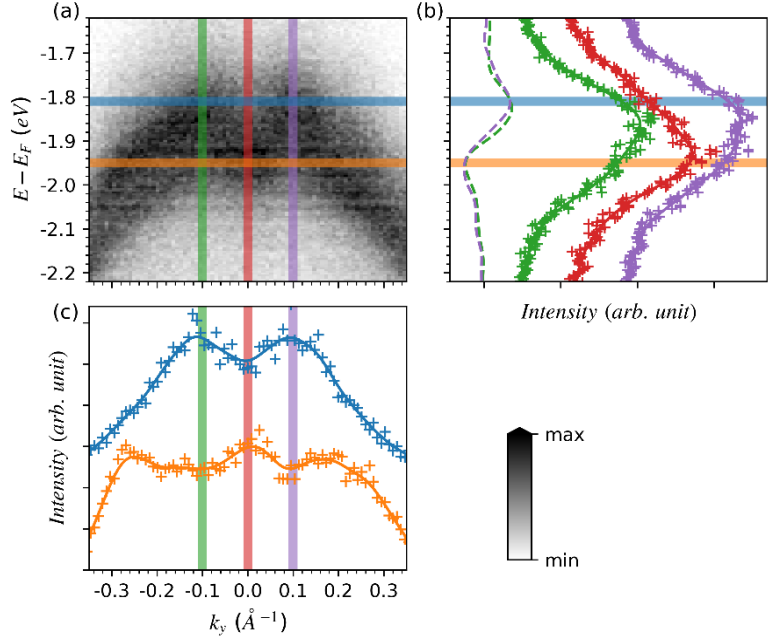

**Figure S7**. Recursive method for the determination of the Rashba parameter. (a) High resolution ARPES image of WS$_{2(1-x)}$Se$_{2x}$ alloy along the ΓM direction. (b) EDCs extracted at Γ and at the two maxima of the blue curve reported in (c), i.e. ky = -0.1, 0, 0.1 Å$^{-1}$ (green, red and violet curves, respectively). (c) MDCs extracted from the maxima (-1.81 ± 0.01 eV) and minima (-1.95 ± 0.01 eV) of the dashed curves in (b), which are the difference between the EDC at Γ (red curve) and the other two EDC (violet and green curves, with the corresponding dashed curves sharing the same line color).

## VIII. Piezoelectric tensors in monolayer WS$_{2(1-x)}$Se$_{2x}$:

The piezoelectric properties of a crystal are described by the piezoelectric tensors defined as:

$$e_{ijk} = \frac{\partial P_i}{\partial \varepsilon_{jk}} = \frac{\partial \sigma_{jk}}{\partial E_i}$$

$$d_{ijk} = \frac{\partial P_i}{\partial \sigma_{jk}} = \frac{\partial \varepsilon_{jk}}{\partial E_i}$$

where $\vec{P}$ is the polarization, $\varepsilon$ the strain, $\sigma$ the stress, $\vec{E}$ the macroscopic electric field and the indices (i, j, k) = {1,2,3} indicate the three spatial directions *x* (armchair), *y* (zig-zag), and *z* (vertical). By using the Voigt notation, the third-rank tensors $e_{ijk}$ and $d_{ijk}$ simplify to $e_{il}$ and $d_{il}$ with $l = \{1, \dots, 6\}$. However, the number of independent, non-null components of $e_{il}$ and $d_{il}$ is determined by the geometry and symmetry of the system. In 2D materials such as TMDs, the vertical direction (*z*-axis) is stress free implying that the stress and strain components are non-null only in the *x-y* plane. Moreover, by assuming a $D_{3h}$ symmetry, the monolayer WS$_2$ possess a vertical symmetry with respect to the atomic plane where the W atoms are located, as well as a lateral symmetry with respect to the *y*-axis (armchair direction). This implies that the piezoelectric tensor and the piezoelectric strain tensor simply read:

$$e_{il} = \begin{pmatrix} e_{11} & -e_{11} & 0 \\ 0 & 0 & -e_{11}/2 \\ 0 & 0 & 0 \end{pmatrix}$$

$$d_{il} = \begin{pmatrix} d_{11} & -d_{11} & 0 \\ 0 & 0 & -2d_{11} \\ 0 & 0 & 0 \end{pmatrix}$$

Such equations say that a finite polarization could be only along the armchair direction.

In the monolayer composed of $WS_{2(1-x)}Se_{2x}$ alloy, however, the random substitution of S atoms with Se produces an inhomogeneous charge distribution also along the zig-zag direction and hence a breaking of the crystal symmetry along the *y*-axis, making non-null the corresponding elements of the piezoelectric tensor. This corroborates the existence of a finite unidirectional polarization along the zig-zag direction (oriented along the ΓK in the reciprocal space) of only the monolayer $WS_{2(1-x)}Se_{2x}$ flakes and the consequent band splitting along the Γ-M direction.